\documentclass[12pt]{article}

\usepackage{epsf,amsfonts,hyperref}
\renewcommand{\appendix}[1]{
    \addtocounter{section}{1}
    \setcounter{equation}{0}
    \renewcommand{\thesection}{\Alph{section}}
    \section*{Appendix \thesection\protect\indent #1}
    \addcontentsline{toc}{section}{Appendix \thesection\ \ \ #1}
}
\newcommand\encadremath[1]{\vbox{\hrule\hbox{\vrule\kern8pt
\vbox{\kern8pt \hbox{$\displaystyle #1$}\kern8pt}
\kern8pt\vrule}\hrule}}
\def\enca#1{\vbox{\hrule\hbox{
\vrule\kern8pt\vbox{\kern8pt \hbox{$\displaystyle #1$}
\kern8pt} \kern8pt\vrule}\hrule}}

\newcommand\figureframex[3]{
\begin{figure}[bth]
\hrule\hbox{\vrule\kern8pt
\vbox{\kern8pt \vbox{
\begin{center}
{\mbox{\epsfxsize=#1.truecm\epsfbox{#2}}}
\end{center}
\caption{#3}
}\kern8pt}
\kern8pt\vrule}\hrule
\end{figure}
}
\newcommand\figureframey[3]{
\begin{figure}[bth]
\hrule\hbox{\vrule\kern8pt
\vbox{\kern8pt \vbox{
\begin{center}
{\mbox{\epsfysize=#1.truecm\epsfbox{#2}}}
\end{center}
\caption{#3}
}\kern8pt}
\kern8pt\vrule}\hrule
\end{figure}
}

\renewcommand{\thesection}{\arabic{section}}

\makeatletter
\@addtoreset{equation}{section}
\makeatother
\newtheorem{theorem}{Theorem}[section]

\newtheorem{remark}{Remark}[section]
\newtheorem{proposition}{Proposition}[section]
\newtheorem{lemma}{Lemma}[section]
\newtheorem{corollary}{Corollary}[section]
\newtheorem{definition}{Definition}[section]
\def\br{\begin{remark}\rm\small}
\def\er{\end{remark}}
\def\bt{\begin{theorem}}
\def\et{\end{theorem}}
\def\bd{\begin{definition}}
\def\ed{\end{definition}}
\def\bp{\begin{proposition}}
\def\ep{\end{proposition}}
\def\bl{\begin{lemma}}
\def\el{\end{lemma}}
\def\bc{\begin{corollary}}
\def\ec{\end{corollary}}
\def\beaq{\begin{eqnarray}}
\def\eeaq{\end{eqnarray}}
\newcommand{\proof}[1]{{\noindent \bf proof:}\par
{#1} $\square$
\bigskip}

\newcommand{\eq}[1]{Eq.~(\ref{#1})}

\newcommand{\beq}{\begin{equation}}
\newcommand{\eeq}{\end{equation}}
\newcommand{\bea}{\begin{eqnarray}}
\newcommand{\eea}{\end{eqnarray}}

\renewcommand{\and}{{\qquad {\rm and} \qquad}}

\newcommand{\virg}{{\qquad , \qquad}}

 \newcommand{\Tr}{{\,\rm Tr}\:}

\newcommand{\Res}{\mathop{\,\rm Res\,}}

\newcommand{\td}[1]{{\tilde{#1}}}

\newcommand{\ee}[1]{{{\rm e}^{#1}}}

\newcommand{\Pint}{{\int\kern -1.em -\kern-.25em}}

\renewcommand{\L}{\Lambda}

\newcommand{\curve}{{\cal E}}

\begin{document}
%=============================Page de titre===============
%\date{??}
%\author{Eynard}
%\title{Correlation functions for hermitian random matrices}
%\topmargin .5cm \textheight 21.5cm \textwidth 15.8cm
%\oddsidemargin 0.54cm
%\evensidemargin 0.54cm
\sloppy

%\maketitle

\pagestyle{empty}
\hfill SPT-07/065
\addtolength{\baselineskip}{0.20\baselineskip}
\begin{center}
\vspace{26pt}
{\large \bf {Weil-Petersson volume of moduli spaces, Mirzakhani's recursion and matrix models}}
\newline
\vspace{26pt}

{\sl B.\ Eynard}\hspace*{0.05cm}\footnote{ E-mail: bertrand.eynard@cea.fr },
 {\sl N.\ Orantin}\hspace*{0.05cm}\footnote{ E-mail: orantin@cea.fr }\\
\vspace{6pt}
Service de Physique Th\'{e}orique de Saclay,\\
F-91191 Gif-sur-Yvette Cedex, France.\\
\end{center}

\vspace{20pt}
\begin{center}
{\bf Abstract}

We prove that Mirzakhani's recursions for the volumes of moduli space of Riemann surfaces are a special case of random matrix recursion relations, and therefore we confirm again that Kontsevich's integral is a generating function for those volumes.
As an application, we propose a formula for the Weil-Petersson volume Vol$({\cal M}_{g,0})$.

\end{center}
%-----------------------------ABSTRACT--------------------------------------
%
%Abstract

%\begin{center}

%\end{center}

%\newpage
%\pagestyle{empty}

%\section*{}

%\newpage
\vspace{26pt}
\pagestyle{plain}
\setcounter{page}{1}

%*********************************************************************
%==================== ARTICLE ========================================
%*********************************************************************

\section{Introduction}

Let
\bea
V_{g,n}(L_1,\dots,L_n)
&=& {\rm Vol}({\cal M}_{g,n}) \cr
&=&\sum_{d_0+\dots+d_n=3g-3+n} \left(\prod_{i=0}^n {1\over d_i!}\right) \,\, \left<\kappa_1^{d_0} \tau_{d_1}\dots \tau_{d_n}\right>_{g,n}\,\, L_1^{2d_1}\dots L_n^{2d_n} \cr
\eea
denote the volume of the moduli space of curves of genus $g$, with $n$ geodesic boundaries of lengths $L_1,\dots,L_n$, measured with the Weil-Petersson metrics.
Using Teichmuller pants decomposition and hyperbolic geometry, M. Mirzakhani \cite{Mirza1} has found a recursion relation among the $V_{g,n}$'s, which allows to compute all of them in a recursive manner. It was then observed \cite{mulsaf} that this recursion relation is equivalent to Virasoro constraints.

In fact, Mirzakhani's recursion relation takes a form \cite{Xu} which is amazingly similar to the recursion relation obeyed by matrix models correlation functions (\cite{eynloop1mat,EOFg}) and which were indeed initialy derived from loop equations \cite{eynloop1mat}, i.e. Virasoro constraints.

Here we make this observation more precise, and we prove that after Laplace transform, Mirzakhani's recursion is {\bf identical} to the recursion of \cite{EOFg} for 
the Kontsevich integral with times (Kontsevich's integral depends only on odd times):
\beq
Z(t_k) = \int dM \,\ee{-N\Tr [{M^3\over 3}+\L M^2]}
\virg
t_{2k+3} = {1\over N}\Tr \L^{-(2k+3)} = \,{(2\pi)^{2k}\,\,(-1)^k\,\over (2k+1)!}+2\delta_{k,0}
.
\eeq

%the spectral curve:
%\beq
%y = -\,{1\over 4\pi}\,\sin{(2\pi\sqrt{x}\,)}
%\eeq
%so that
%\beq
%y(x) = \sqrt{x} - {1\over 2}\sum_{k=0}^\infty t_{2k+3}\,x^{k+1/2}
%\eeq
%we prove that
%\beq
%W_1(x) = \left< {1\over N} \,\Tr {\ee{i\pi \L}\over x-\L^2}\, M \right>
%\eeq

\section{Laplace transform}

Define the Laplace transforms of the $V_{g,n}$'s:
\bea
&& W^g_{n}(z_1,\dots,z_n) \cr
&=& 2^{-m_{g,n}}\,\int_{0}^\infty dL_1\dots dL_n \ee{-\sum_i z_i L_i}\,\, \prod_{i=1}^n L_i \,\,\, V_{g,n}(L_1,\dots,L_n) \cr
&=& 2^{-m_{g,n}}\,{\displaystyle \sum_{d_0+\dots+d_n=3g-3+n}} \left(\prod_{i=0}^n {1\over d_i!}\right) \,\, \left<\kappa_1^{d_0} \tau_{d_1}\dots \tau_{d_n}\right>_{g,n}\,\, {(2d_1+1)!\over z_1^{2d_1+2}}\dots {(2d_n+1)!\over z_n^{2d_n+2}} \cr
\eea
where (see \cite{Mirza1}) $m_{g,n}=\delta_{g,1}\delta_{n,1}$.

Since the $V_{g,n}$'s are even polynomials of the $L_i$'s, of degree $2d_{g,n}$ where 
\beq
d_{g,n}={\rm dim}\, {\cal M}_{g,n}=3g-3+n
\eeq
the $W^g_n$'s are even polynomials of the $1/z_i$'s of degree $2d_{g,n}+2$.
Let us also define:
\beq
W_1^0 = 0
\eeq
\beq
W_2^0(z_1,z_2) = {1\over (z_1-z_2)^2}
\eeq
and
\beq
dE_u(z) = {1\over 2}\left( {1\over z-u}-{1\over z+u}\right)
.
\eeq

We prove the following theorems:
\bt
For any $2g-2+n+1>0$, the $W_{n+1}^g$ satisfy the recursion relation
\beq\label{eqrecW}
\encadremath{
\begin{array}{rcl}
W^g_{n+1}(z,K)
&=&  \Res_{u\to 0} {\pi dE_u(z)\over u\sin{2\pi u}} \left[\sum_{h=0}^g\sum_{J\subset K}\, W^h_{1+|J|}(u,J) W^{g-h}_{1+n-|J|}(-u,K/J) \right.\cr
&& \qquad \qquad \qquad \qquad \left. + W^{g-1}_{n+2}(u,-u,K) \right]
\cr
\end{array}}
\eeq
where the RHS includes all possible $W_k^{h}$, including $W_1^0=0$ and $W_2^0$,
and where
\beq
K=\{ z_1,\dots, z_n \}
\eeq
is a set of $n$ variables.
\et

\proof{This relation is merely the Laplace transform of Mirzakhani's recursion. See the appendix  for a detailed proof.}

\bc
$W_n^g$ are the invariants defined in \cite{EOFg} for the curve:
\beq
\left\{
\begin{array}{l}
x(z) = z^2 \cr
-2y(z) = {\sin{(2\pi z)}\over 2\pi} = z  -  2{\pi^2\over 3}z^3 + {2\pi^4\over 15}z^5 - {4\pi^6\over 315}z^7 + {2\pi^8\over 2835} z^9 + \dots
\end{array}
\right.
\eeq
which is a special case of Kontsevich's curve:
\beq
Z(t_k) = \int dM \,\ee{-N\Tr [{M^3\over 3}+\L M^2]}
\virg
t_k = {1\over N}\Tr \L^{-k} = {(2\pi)^{k-3}\,\,\sin{(\pi k/2)}\,\over (k-2)!}
\eeq
For instance we have:
\beq
\ln{Z(t_k)} = \sum_{g=0}^\infty N^{2-2g} W_0^g
\eeq
($W_0^g$ is often noted $-F_g$ in the litterature).
\ec

\proof{Eq. \ref{eqrecW} is precisely the definiton of the invariants of \cite{EOFg} for the curve
\beq
\left\{
\begin{array}{l}
x(z) = z^2 \cr
-2y(z) = {\sin{(2\pi z)}\over 2\pi} = z  -  2{\pi^2\over 3}z^3 + {2\pi^4\over 15}z^5 - {4\pi^6\over 315}z^7 + {2\pi^8\over 2835} z^9 + \dots
\end{array}
\right.
\eeq
And it was proved in \cite{EOFg} that this curve is a special case of Kontsevich's curve:
\beq
\left\{
\begin{array}{l}
x(z) = z^2 \cr
y(z) = z-{1\over 2}\sum_{j=0}^\infty t_{j+2} z^j
\end{array}
\right.
\eeq
which corresponds to the computation of the topological expansion of the Kontsevich integral:
\beq
Z(t_k) = \int dM \,\ee{-N\Tr [{M^3\over 3}+\L M^2]}
\virg
t_k = {1\over N}\Tr \L^{-k}
\eeq
\beq
\ln{Z(t_k)} = - \sum_{g=0}^\infty N^{2-2g} F_g
\eeq

}

\bt
For any $2g-2+n>0$ we have:
\beq
(2g-2+n)\, W^g_{n}(K) =  {1\over 4\pi^2}\,\Res_{u\to 0}  \left(u \cos{(2\pi u)} - {1\over 2\pi}\sin{(2\pi u)}\right)    \,\, W^g_{n+1}(u,K)
\eeq
or in inverse Laplace transform:
\beq
\encadremath{
(2g-2+n)\, V_{g,n}(K) =  {1\over 2i\pi}\, V'_{g,n+1}(K,2i\pi)
}\eeq
where $'$ means the derivative with respect to the $n+1^{\rm th}$ variable.
\et

\proof{This is a mere application of theorem 4.7. in \cite{EOFg}, as well as its Laplace transform.}

In particular with $n=0$ we get:
\beq
V_{g,0} = {\rm Vol}({\cal M}_{g,0}) = {1\over 2g-2}\,\,{V'_{g,1}(2i\pi)\over 2i\pi}
\eeq
for instance for $g=2$:
\beq
V_{2,0} = {43 \pi^6\over 2160}
.
\eeq

\subsection{Examples}

From \cite{Mirza1} we get:
\beq
W^0_{3} = {1\over z_1^2 z_2^2 z_3^2}
\eeq
\beq
W^1_{1} = {1\over 8 z_1^4} + {\pi^2\over 12 z_1^2}
\eeq
\beq
W^0_{4} = {1\over z_1^2 z_2^2 z_3^2 z_4^2} \left(2\pi^2 + 3({1\over z_1^2}+{1\over z_2^2}+{1\over z_3^2}+{1\over z_4^2})\right)
\eeq
\beq
W^1_{2} = {1\over z_1^2 z_2^2 } \left( { \pi^4\over 4} + {\pi^2\over 2} ({1\over z_1^2}+{1\over z_2^2}) + {5\over 8 z_1^4}+{5\over 8 z_2^4}+{3\over 8 z_1^2 z_2^2}  \right)
\eeq
\beq
W^0_{5} = {1\over z_1^2 z_2^2 z_3^2 z_4^2 z_5^2} \left(10\pi^4 + 18\pi^2 \sum_i {1\over z_i^2}+ 15\sum_i {1\over z_i^4}  + 18 \sum_{i<j} {1\over z_i^2 z_j^2} \right)
\eeq
\beq
W^2_{1} = {1\over 192 z_1^2}\left( 29 \pi^8 + {338\pi^6\over 5 z_1^2}
+{139 \pi^4 \over z_1^4} + {203 \pi^2\over z_1^6} + {315\over 2 z_1^8} \right)
\eeq

%\bigskip

%On the other hand, from \cite{EOFg}, we find that Kontsevich's integral correlation functions are:
%\beq
%W^0_{3} = {1\over t_3-2}\,{1\over z_1^2 z_2^2 z_3^2}
%\eeq
%\beq
%W^1_{1} = {1\over 8 (t_3-2)\, z_1^4} - {t_5\over 8(t_3-2)^2 z_1^2}
%\eeq
%\beq
%W^0_{4} = {3\over (t_3-2)^2\, z_1^2 z_2^2 z_3^2 z_4^2} \left({1\over z_1^2}+{1\over z_2^2}+{1\over z_3^2}+{1\over z_4^2} - {t_5\over t_3-2} \right)
%\eeq
%etc...

%\bigskip

Those functions are the same as those which appear in section 10.4.1 of \cite{EOFg}, for the Kontsevich curve with times:
\beq
t_3-2=1, t_5=-{2\pi^2\over 3}, t_7= {2\pi^4\over 15}, t_9=-{4\pi^6\over315}, t_{11}={2\pi^8\over 2835}, \dots
\eeq
i.e. the rational curve:
\beq
{\curve}_{K} =
\left\{
\begin{array}{l}
x(z) = z^2 \cr
-2y(z) = {\sin{(2\pi z)}\over 2\pi} = z  -  2{\pi^2\over 3}z^3 + {2\pi^4\over 15}z^5 - {4\pi^6\over 315}z^7 + {2\pi^8\over 2835} z^9 + \dots
\end{array}
\right.
\eeq
It is to be noted that those $t_k$'s are closely related to the $\beta_k$'s of \cite{mulsaf,Xu}.

\section{Conclusion}

We have shown that, after Laplace transform, Mirzakhani's recursions are nothing but the solution of loop equations (i.e. Virasoro constraints) for the Kontsevich integral with some given set of times.
It would be interesting to understand what the invariants of \cite{EOFg} compute for an arbitrary spectral curve (for instance for other Kontsevich times).

\subsection*{Acknowledgements}

This work is partly supported by the Enigma European network MRT-CT-2004-5652, by the ANR project G\'eom\'etrie et int\'egrabilit\'e en physique math\'ematique ANR-05-BLAN-0029-01,
 by the Enrage European network MRTN-CT-2004-005616,
 by the European Science foundation through the Misgam program,
 by the French and Japaneese governments through PAI Sakura, by the Quebec government with the FQRNT.

\setcounter{section}{0}

\appendix{Laplace transform of the equations}
\label{appendix1}

Let us write:
\beq
L_K=\{L_1,\dots,L_n\}
\eeq
\beq
H_n^g(x,y,L_K) = xy V_{g-1,n+2}(x,y,L_K) + \sum_{h=0}^{g}\sum_{J\in K} x V_{h,1+|J|}(x,L_J) y V_{g-h,n+1-|J|}(y,L_{K/J})
\eeq
where all the $V_{h,k}$ terms in the RHS are such that $2h+k-2>0$ (i.e. stable curves only), as well as their
laplace transform:
\beq
\widetilde{H}_n^{(g)}(z,z',L_K):= \int_0^\infty dx \int_0^\infty dy e^{-zx} e^{-z'y} H_n^g(x,y,L_K).
\eeq

Mirzakhani's recursion reads:
\beq\label{apprec}
\begin{array}{rcl}
2 L V_{g,n+1}(L,L_K) &=&  {\displaystyle \int_{0}^L dt \int_0^\infty dx \int_0^\infty dy K(x+y,t) H^g_n(x,y,L_K) }\cr
&+& {\displaystyle \sum_{m=1}^n \int_0^L dt \int_0^\infty dx (K(x,t+L_m)+K(x,t-L_m)) x V_{g,n-1}(x,\hat{L}_m) }\cr
\end{array}
\eeq
where
\beq
K(x,t) = {1\over 1+\ee{\left({x+t\over 2}\right)}} + {1\over 1+\ee{\left({x-t\over 2}\right)}}
\eeq
and $\hat{L}_m=L_K/\{L_m\}$.

Let $\td{H}_n^g$  be the Laplace transform of $H_n^g$ with respect to $x$ and $y$.

The Laplace transform of the first term in eq.\ref{apprec} is:
\bea
&& \sum_{\epsilon=\pm 1}\int_0^\infty dL\, \ee{-zL} \int_0^L dt \int_0^\infty dx \int_0^\infty dy {1\over 1+\ee{{x+y+\epsilon t\over 2}}} \,\,H_n^g(x,y,L_K) \cr
&=& \sum_{\epsilon=\pm 1} \int_0^\infty dt \int_t^\infty dL\, \ee{-zL}  \int_0^\infty dx \int_0^\infty dy {1\over 1+\ee{{x+y+\epsilon t\over 2}}} \,\,H_n^g(x,y,L_K) \cr
&=& \sum_{\epsilon=\pm 1} {1\over z} \int_0^\infty dt\,  \ee{-zt}  \int_0^\infty dx \int_0^\infty dy {1\over 1+\ee{{x+y+\epsilon t\over 2}}} \,\,H_n^g(x,y,L_K) \cr
&=& -\sum_{j=1}^\infty {1\over z} \int_0^\infty dt\,  \ee{-zt}  \int_0^\infty dx \int_0^\infty dy (-1)^j \ee{-{j\over2}(x+y+t)} \,\,H_n^g(x,y,L_K) \cr
&& + \sum_{j=0}^\infty  {1\over z} \int_0^\infty dx \int_0^\infty dy \int_{x+y}^\infty dt \, \ee{-zt}  (-1)^j \ee{{j\over2}(x+y-t)} \,\,H_n^g(x,y,L_K) \cr
&& - \sum_{j=1}^\infty  {1\over z} \int_0^\infty dx \int_0^\infty dy \int_0^{x+y} dt \, \ee{-zt} (-1)^j \ee{-{j\over2}(x+y-t)} \,\,H_n^g(x,y,L_K) \cr
&=& -\sum_{j=1}^\infty {1\over z}  \int_0^\infty dx \int_0^\infty dy {(-1)^j\over z+{j\over 2}} \ee{-{j\over 2}(x+y)} \,\,H_n^g(x,y,L_K) \cr
&& + \sum_{j=0}^\infty  {1\over z} \int_0^\infty dx \int_0^\infty dy   {(-1)^j\over z+{j\over 2}} \ee{-z(x+y)} \,\,H_n^g(x,y,L_K) \cr
&& - \sum_{j=1}^\infty  {1\over z} \int_0^\infty dx \int_0^\infty dy  {(-1)^j\over z-{j\over 2}}\,(1-\ee{-(z-{j\over 2})(x+y)}) \ee{-{j\over 2}(x+y)} \,\,H_n^g(x,y,L_K) \cr
%&=& -\sum_{j=1}^\infty {1\over z}  \int_0^\infty dx \int_0^\infty dy {(-1)^j\over z+j} \ee{-j(x+y)} \,\,H_n^g(x,y,L_K) \cr
%&& - \sum_{j=1}^\infty  {1\over z} \int_0^\infty dx \int_0^\infty dy  {(-1)^j\over z-j}\,\ee{-j(x+y)} \,\,H_n^g(x,y,L_K) \cr
%&& + \sum_{j=0}^\infty  {1\over z} \int_0^\infty dx \int_0^\infty dy   {(-1)^j\over z+j} \ee{-z(x+y)} \,\,H_n^g(x,y,L_K) \cr
%&& + \sum_{j=1}^\infty  {1\over z} \int_0^\infty dx \int_0^\infty dy  {(-1)^j\over z-j}\,\ee{-z(x+y)}  \,\,H_n^g(x,y,L_K) \cr
&=& - 2 \sum_{j=1}^\infty   {(-1)^j\over z^2-\left(j\over 2\right)^2} \,\,\td{H}_n^g({j\over2},{j\over2},L_K) +  {1\over z^2}  \,\, \td{H}_n^g(z,z,L_K)  \cr
&& + 2\sum_{j=1}^\infty   {(-1)^j\over z^2-\left(j\over 2\right)^2}  \,\,\td{H}_n^g(z,z,L_K) \cr
&=& - 2 \sum_{j=1}^\infty   {(-1)^j\over z^2-\left(j\over 2\right)^2} \,\,\td{H}_n^g({j\over2},{j\over2},L_K)  +  {2 \pi\over z \sin{2 \pi z}}\,   \,\, \td{H}_n^g(z,z,L_K)  \cr
&=& \left(\Res_{u\to z} + \sum_{j=1}^\infty \Res_{u\to \pm {j\over 2}} \right) {du\over u-z}\, {2 \pi\over u\sin{(2 \pi u)}}  \,\, \td{H}_n^g(u,u,L_K)  \cr
&=& \Res_{u\to 0} {du\over z-u} \,{2 \pi\over u\sin{(2 \pi u)}}  \,\, \td{H}_n^g(u,u,L_K)  \cr
&=& \Res_{u\to 0} {2 \pi \, du\over u\sin{(2 \pi u)}} \, dE_u(z) \,\, \td{H}_n^g(u,u,L_K)  \cr
\eea

Using the notation
\beq
R(x,t,L_m):= (K(x,t+L_m)+K(x,t-L_m)),
\eeq
the Laplace transform of the second term in eq.\ref{apprec} is:
\bea
&& \int_0^\infty dL_m \ee{-z_m L_m}\, \int_0^\infty dL\, \ee{-zL}\, \int_0^L dt \int_0^\infty dx R(x,t,L_m) x V_{g,n-1}(x,\hat{L}_m) \cr
&=& {1\over z}\,\int_0^\infty dx \,\int_0^\infty dL_m \ee{-z_m L_m}\, \int_0^\infty dt\, \ee{-zt}\, R(x,t,L_m) x V_{g,n-1}(x,\hat{L}_m) \cr
&=& {1\over z}\,\int_0^\infty dx \,\int_0^\infty dL_m \ee{-z_m L_m}\, \int_{L_m}^\infty dt\, \ee{-z(t-L_m)}\,  K(x,t) x V_{g,n-1}(x,\hat{L}_m) \cr
&& + {1\over z}\,\int_0^\infty dx \,\int_0^\infty dL_m \ee{-z_m L_m}\, \int_{-L_m}^\infty dt\, \ee{-z(t+L_m)}\,  K(x,t) x V_{g,n-1}(x,\hat{L}_m) \cr
&=& {1\over z}\,\int_0^\infty dx \,\int_{0}^\infty dt\, \ee{-zt}\, \int_0^t dL_m \ee{-(z_m-z) L_m}\,  K(x,t) x V_{g,n-1}(x,\hat{L}_m) \cr
&& + {1\over z}\,\int_0^\infty dx \,\int_{0}^\infty dt\, \ee{-zt}\, \int_0^\infty dL_m \ee{-(z_m+z) L_m}\,  K(x,t) x V_{g,n-1}(x,\hat{L}_m) \cr
&& + {1\over z}\,\int_0^\infty dx \,\int_{-\infty}^0 dt\, \ee{-zt}\, \int_{-t}^\infty dL_m \ee{-(z_m+z) L_m}\,  K(x,t) x V_{g,n-1}(x,\hat{L}_m) \cr
&=& {1\over z}\,\int_0^\infty dx \,\int_{0}^\infty dt\,  {\ee{-zt}-\ee{-z_m t}\over z_m-z}\,  K(x,t) x V_{g,n-1}(x,\hat{L}_m) \cr
&& + {1\over z}\,\int_0^\infty dx \,\int_{0}^\infty dt\, {\ee{-zt}+\ee{z_m t}\over z_m+z}\,  K(x,t) x V_{g,n-1}(x,\hat{L}_m) \cr
&& + {1\over z}\,\int_0^\infty dx \,\int_0^{\infty} dt\,  {\ee{- z_m t}\over z_m+z}\,  K(x,t) x V_{g,n-1}(x,\hat{L}_m) \cr
&=& {1\over z}\,\int_0^\infty dx \,\int_{0}^\infty dt\,  \left({\ee{-zt}-\ee{-z_m t}\over z_m-z}+{\ee{-zt}+\ee{-z_m t}\over z_m+z}\right)\,  K(x,t) x V_{g,n-1}(x,\hat{L}_m) \cr
&=& {1\over z}\,\int_0^\infty dx \,\int_{0}^\infty dt\,  {2 z_m \ee{-zt} -2z \ee{-z_m t} \over (z_m^2-z^2)}\,  {1\over 1+\ee{{x+t\over 2}}}\, x V_{g,n-1}(x,\hat{L}_m) \cr
&& +{1\over z}\,\int_0^\infty dx \,\int_{0}^x dt\,  {2 z_m \ee{-zt} -2z \ee{-z_m t} \over (z_m^2-z^2)}\,  {1\over 1+\ee{{x-t\over 2}}}\, x V_{g,n-1}(x,\hat{L}_m) \cr
&& +{1\over z}\,\int_0^\infty dx \,\int_{x}^\infty dt\,  {2 z_m \ee{-zt} -2z \ee{-z_m t} \over (z_m^2-z^2)}\,  {1\over 1+\ee{{x-t\over 2}}}\, x V_{g,n-1}(x,\hat{L}_m) \cr
&=& - \sum_{j=1}^\infty {(-1)^j\over z}\,\int_0^\infty dx \,\int_{0}^\infty dt\,  {2 z_m \ee{-zt} -2z \ee{-z_m t} \over (z_m^2-z^2)}\,  \ee{-{j\over2}(x+t)}\, x V_{g,n-1}(x,\hat{L}_m) \cr
&& - \sum_{j=1}^\infty {(-1)^j\over z}\,\int_0^\infty dx \,\int_{0}^x dt\,  {2 z_m \ee{-zt} -2z \ee{-z_m t} \over (z_m^2-z^2)}\,  \ee{-{j \over 2}(x-t)}\, x V_{g,n-1}(x,\hat{L}_m) \cr
&& + \sum_{j=0}^\infty {(-1)^j\over z}\,\int_0^\infty dx \,\int_{x}^\infty dt\,  {2 z_m \ee{-zt} -2z \ee{-z_m t} \over (z_m^2-z^2)}\,  \ee{{j \over 2}(x-t)}\, x V_{g,n-1}(x,\hat{L}_m) \cr
&=& - \sum_{j=1}^\infty {(-1)^j\over z}\,\int_0^\infty dx \, { {2 z_m\over z+{j\over 2}} - {2z\over z_m+{j\over2}} \over (z_m^2-z^2)}\,   \ee{-{j\over 2}x} \, x V_{g,n-1}(x,\hat{L}_m) \cr
&& - \sum_{j=1}^\infty {(-1)^j\over z}\,\int_0^\infty dx \,\int_{0}^x dt\,  {2 z_m\,{\ee{-{j\over2} x}-\ee{-zx}\over z-{j\over 2}}  -2z\,{\ee{-{j\over 2}x}-\ee{-z_m x}\over z_m-{j\over2}}  \over (z_m^2-z^2)}\, \, x V_{g,n-1}(x,\hat{L}_m) \cr
&& + \sum_{j=0}^\infty {(-1)^j\over z}\,\int_0^\infty dx \,\int_{x}^\infty dt\,  {{2 z_m \ee{-zx}\over z+{j\over2}} -{2z \ee{-z_m x}\over z_m+{j\over2}} \over (z_m^2-z^2)}\, \, x V_{g,n-1}(x,\hat{L}_m) \cr
&=& - 2 \sum_{j=1}^\infty {(-1)^j\over z}\, \, { z+z_m+{j\over2} \over (z_m+z)(z+{j\over2})(z_m+{j\over2})}\,   W_{g,n-1}({j\over2},\hat{L}_m) \cr
&& - 2\sum_{j=1}^\infty {(-1)^j\over z}\, \, { z+z_m-{j\over2} \over (z_m+z)(z-{j\over2})(z_m-{j\over2})}\, \, W_{g,n-1}({j\over2},\hat{L}_m) \cr
&& + 2\sum_{j=1}^\infty {(-1)^j\over z}\, \,  { z_m \over (z-{j\over2}) (z_m^2-z^2)}\, \, W_{g,n-1}(z,\hat{L}_m) \cr
&& - 2\sum_{j=1}^\infty {(-1)^j}\, \,     \,{1\over (z_m-{j\over2})(z_m^2-z^2)}\, \, W_{g,n-1}(z_m,\hat{L}_m) \cr
&& + 2\sum_{j=0}^\infty {(-1)^j\over z}\, \,  { z_m \over (z+{j\over2})(z_m^2-z^2)}\, \, W_{g,n-1}(z,\hat{L}_m) \cr
&& - 2\sum_{j=0}^\infty {(-1)^j}\, \,  { 1 \over (z_m+{j\over2})(z_m^2-z^2)}\, \, W_{g,n-1}(z_m,\hat{L}_m) \cr
&=& - 4 \sum_{j=1}^\infty \Res_{u\to \pm j} {\pi du\over \sin{2 \pi u}}\,\, {1\over z}\, \, { z+z_m+u \over (z_m+z)(z+u)(z_m+u)}\,   W_{g,n-1}(u,\hat{L}_m) \cr
&& + 4 \,  { z_m \pi \over z \sin{(2 \pi z)}\, (z_m^2-z^2)}\, \, W_{g,n-1}(z,\hat{L}_m) \cr
&& - 4   \,{\pi \over \sin{(2 \pi z_m)}\,\,(z_m^2-z^2)}\, \, W_{g,n-1}(z_m,\hat{L}_m) \cr
&=& - 4 \sum_{j=1}^\infty \Res_{u\to \pm {j\over 2}} {\pi du\over \sin{2 \pi u}} \, \, {  z_m \over (z^2-u^2)(z_m^2-u^2)}\,   W_{g,n-1}(u,\hat{L}_m) \cr
&& - 4 \Res_{u\to z,z_m} {\pi du\over \sin{2 \pi u}}\,\,  { z_m \over (z_m^2-u^2)(z^2-u^2)}\, \, W_{g,n-1}(u,\hat{L}_m) \cr
&=&  4 \Res_{u\to 0} {\pi du\over \sin{2 \pi u}} \, \, {  z_m \over (z^2-u^2)(z_m^2-u^2)}\,   W_{g,n-1}(u,\hat{L}_m) \cr
&=& 2 \Res_{u\to 0} {\pi du\over 2u \sin{2 \pi u}} \, \, \left({1 \over z-u}-{1\over z+u}\right)\,\,\left({1\over z_m-u}+{1\over z_m+u}\right)\,   W_{g,n-1}(u,\hat{L}_m) \cr
&=& 4 \Res_{u\to 0} {\pi du\over 2u \sin{2 \pi u}} \, \, \left({1 \over z-u}-{1\over z+u}\right)\,\,{1\over z_m-u}\,   W_{g,n-1}(u,\hat{L}_m) \cr
\eea
After taking the derivative with respect to $z_m$ that gives the expected term:
\beq
\Res_{u\to 0} {2 \pi du\over u \sin{2 \pi u}} \, \, dE_u(z)\,\, 2\, W_2^0(u,z_m)   W_{g,n-1}(u,\hat{L}_m)
\eeq
and therefore the Laplace transform of \eq{apprec} gives the relation \eq{eqrecW}.


\begin{thebibliography}{99}

\bibitem{eynloop1mat} B. Eynard, ``Topological expansion for the 1-hermitian matrix model correlation functions'',
JHEP/024A/0904, hep-th/0407261.

\bibitem{EOFg} B.Eynard, N.Orantin,
``Invariants of algebraic curves and topological expansion'', math-ph/0702045.

\bibitem{Xu} K. Liu, H. Xu, ``A simple proof of Mirzakhani's recursion formula of Weil-Petersson volumes'', math.AG/0705.2086.

\bibitem{Mirza1} M. Mirzakhani, ``Simple geodesics and Weil-Petersson volumes of moduli spaces of bordered Riemann surfaces'', Invent. Math. 167, 179-222 (2007).

\bibitem{mulsaf} M. Mulase, B. Safnuk, ``Mirzakhani's recursion relations, Virasoro constraints and the KdV hierarchy'', math.AG/0101147.

\end{thebibliography}
\end{document}